\documentclass[prl,amsmath,twocolumn,amssymb,aps,superscriptaddress,floatfix,preprintnumbers,10pt,notitlepage]{revtex4-1}
\usepackage{times}
\usepackage{graphicx}
\usepackage{amsmath, braket, amsfonts}
\usepackage{amssymb}
\usepackage{natbib}
\usepackage{sidecap}
\usepackage{bm, color, ulem}
\usepackage{comment}

\def\lsim{\lower -0.3ex \hbox{$<$} \kern -0.75em \lower 0.7ex \hbox{$\sim$}}
\def\gsim{\lower -0.3ex \hbox{$>$} \kern -0.75em \lower 0.7ex \hbox{$\sim$}}

\def\sgn{{\rm \sgn}}

\begin{document}
\title{Intrinsic quantized anomalous Hall effect in a moir\'e heterostructure}
\author{M. Serlin}
\thanks{These authors contributed equally}
\affiliation{Department of Physics, University of California, Santa Barbara CA 93106, USA}
\author{C. L. Tschirhart}
\thanks{These authors contributed equally}
\affiliation{Department of Physics, University of California, Santa Barbara CA 93106, USA}
\author{H. Polshyn}
\thanks{These authors contributed equally}
\affiliation{Department of Physics, University of California, Santa Barbara CA 93106, USA}
\author{Y. Zhang}
\affiliation{Department of Physics, University of California, Santa Barbara CA 93106, USA}
\author{J. Zhu}
\affiliation{Department of Physics, University of California, Santa Barbara CA 93106, USA}
\author{K. Watanabe}
\affiliation{National Institute for Materials Science, 1-1 Namiki, Tsukuba 305-0044, Japan}
\author{T. Taniguchi}
\affiliation{National Institute for Materials Science, 1-1 Namiki, Tsukuba 305-0044, Japan}
\author{L. Balents}
\affiliation{Kavli Institute for Theoretical Physics, University of California, Santa Barbara, CA 93106, USA}
\author{A. F. Young}
\email[Electronic address:]{andrea@physics.ucsb.edu}
\affiliation{Department of Physics, University of California, Santa Barbara CA 93106, USA}
\date{\today}

\begin{abstract}
We report the observation of a quantum anomalous Hall effect in twisted bilayer graphene showing Hall resistance quantized to within .1\% of the von Klitzing constant $h/e^2$ at zero magnetic field.
The effect is driven by intrinsic strong correlations, which polarize the electron system into a single  spin and valley resolved moir\'e miniband with Chern number $C=1$.
In contrast to extrinsic, magnetically doped systems, the measured transport energy gap $\Delta/k_B\approx 27$~K is larger than the Curie temperature for magnetic ordering $T_C\approx 9$~K, and Hall quantization persists to temperatures of several Kelvin. Remarkably, we find that electrical currents as small as 1~nA can be used to controllably switch the magnetic order between states of opposite polarization, forming an electrically rewritable magnetic memory.
 \end{abstract}
 \maketitle

Two dimensional insulators can be classified by the topology of their filled energy bands. In the absence of time reversal symmetry, nontrivial band topology manifests experimentally as a quantized Hall conductivity $\sigma_{xy}=C \frac{e^2}{h}$, where $C\neq0$ is the total Chern number of the filled bands. Motivated by fundamental questions about the nature of topological phase transitions\cite{haldane_model_1988} as well as possible applications in resistance metrology\cite{gotz_precision_2018} and topological quantum computing\cite{lian_topological_2018}, significant effort has been devoted to engineering quantum anomalous Hall (QAH) effects showing topologically protected quantized resistance in the absence of an applied magnetic field.
To date, QAH effects have been observed only in a narrow class of materials consisting of transition metal doped (Bi,Sb)$_2$Te$_3$\cite{chang_experimental_2013,chang_high-precision_2015,mogi_magnetic_2015,kou_mapping_2015,kou_scale-invariant_2014,wang_quantum_2013,checkelsky_trajectory_2014}.
In these materials, ordering of the dopant magnetic moments breaks time reversal symmetry, combining with the strongly spin-orbit coupled electronic structure to produce topologically nontrivial Chern bands\cite{yu_quantized_2010}.  However, the performance of these materials is limited by the inhomogeneous distribution of the magnetic dopants, which leads to microscopic structural, charge, and magnetic disorder\cite{lachman_visualization_2015,lee_imaging_2015,wang_visualizing_2015,yasuda_quantized_2017}.   As a result, sub-kelvin temperatures are typically required to observe quantization, despite magnetic ordering temperatures at least one order of magnitude larger.

Moir\'e graphene heterostructures provide the two essential ingredients---topological bands and strong correlations---necessary for engineering intrinsic quantum anomalous Hall effects.
For both graphene on hexagonal boron nitride (hBN) and twisted multilayer graphene, moir\'e patterns generically produce bands with finite Chern number\cite{song_topological_2015,zhang_moire_2018,bultinck_anomalous_2019,zhang_twisted_2019}, with time reversal symmetry of the single particle band structure enforced by the cancellation of Chern numbers in opposite graphene valleys.  In certain heterostructures, notably twisted bilayer graphene (tBLG) with interlayer twist angle $\theta\approx 1.1^\circ$ and rhombohedral graphene aligned to hBN, the bandwidth of these Chern bands can be made exceptionally small\cite{bistritzer_moire_2011,suarez_morell_flat_2010,zhang_moire_2018,chen_evidence_2019}, favoring correlation driven states that break one or more spin, valley, or lattice symmetries.  Experiments have indeed found correlation driven low temperature phases at integer band fillings when these bands are sufficiently flat\cite{cao_correlated_2018,chen_evidence_2019,cao_unconventional_2018,yankowitz_tuning_2019,lu_superconductors_2019}.  Remarkably, states showing magnetic hysteresis indicative of time-reversal symmetry breaking have recently been reported in both tBLG\cite{sharpe_emergent_2019} and ABC/hBN heterostructures\cite{chen_tunable_2019} at commensurate filling.  These systems show large anomalous Hall effects highly suggestive of an incipient Chern insulator at $B=0$.

\begin{figure*}[ht!]
\includegraphics[width= 7.25 in]{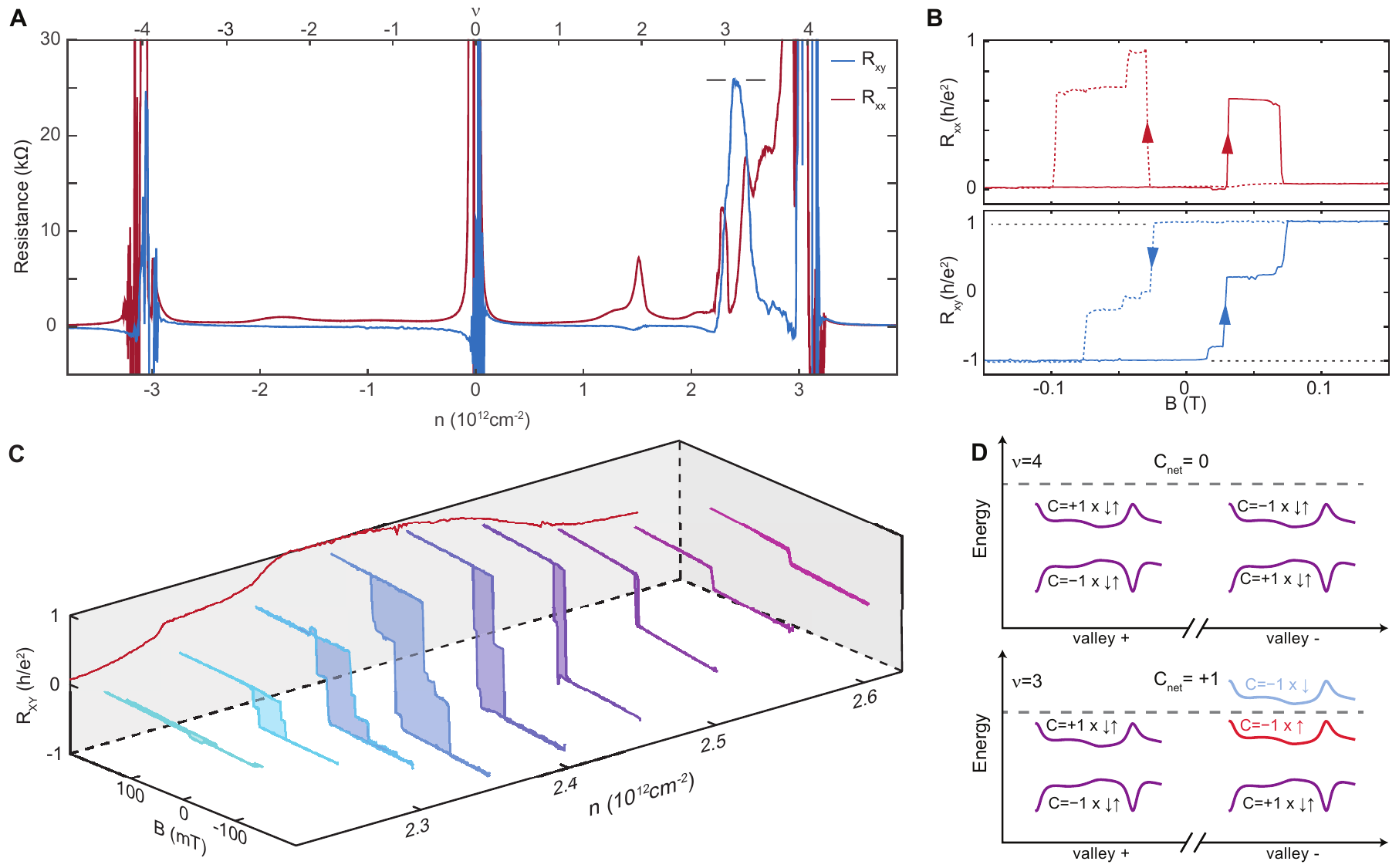}
 \caption{\textbf{Quantized anomalous Hall effect in twisted bilayer graphene}
 (\textbf{A}) Longitudinal resistance $R_{xx}$ and Hall resistance $R_{xy}$ as a function of carrier density $n$ at 150 mT. $R_{xy}$ reaches $h/e^2$ and  $R_{xx}$ approaches zero near $\nu = 3$. Data are corrected for mixing of $R_{xx}$ and $R_{xy}$ components by symmetrizing with respect to magnetic field at $B=\pm150$ mT  \cite{supplementary_information}.
 (\textbf{B}) Longitudinal resistance $R_{xx}$ and Hall resistance $R_{xy}$ measured at $n = 2.37 \times 10^{12} \mathrm{ cm}^{-2}$ as a function of $B$.  Data are corrected for mixing using contact symmetrization\cite{supplementary_information}. Sweep directions are indicated by arrows.
 (\textbf{C}) Hall resistance $R_{xy}$ as a function of magnetic field $B$ and density $n$. Hysteresis loop areas are shaded for clarity. The rear wall shows field-training symmetrized values of $R_{xy}$ at $B = 0$.  $R_{xy}(0)$ becomes nonzero when ferromagnetism appears, and reaches a plateau of $h/e^2$ near a density of $n = 2.37 \times 10^{12} \mathrm{ cm}^{-2}$.
 (\textbf{D}) Schematic band structure at full filling of a moir\'e unit cell ($\nu=4$) and $\nu = 3$.  The net Chern number $C_{net}\neq0$ at $\nu=3$. }
\label{fig:1}
\end{figure*}

Here we report the observation of a QAH effect showing robust zero magnetic field quantization in a flat band ($\theta\approx 1.15\pm0.01^\circ$) tBLG sample.
The electronic structure of flat-band tBLG is described by two distinct bands per spin and valley projection isolated from higher energy dispersive bands by an energy gap.  The total capacity of the flat bands is eight electrons per unit cell,  spanning $-4<\nu<4$, where we define the band filling factor $\nu=n A_m$ with $n$ the electron density and $A_m\approx 130$~$ \mathrm{nm}^2$ the area of the moir\'e unit cell.
Figure \ref{fig:1}A shows the longitudinal and Hall resistances ($R_{xx}$ and $R_{xy}$) measured at a magnetic field $B=150$~mT~\cite{supplementary_information} and temperature $T=1.6$~K as a function of charge density over the entire flat band.  The sample is insulating at the overall charge neutral point and shows a weak resistance peak at $\nu=2$.  In addition, we observe $R_{xy}$ approaching $h/e^2$ in a narrow range of density near $\nu=3$, concomitant with a deep minimum in $R_{xx}$ reminiscent of an integer quantum Hall state.

Figure \ref{fig:1}B shows the magnetic field dependence of both $R_{xx}$ and $R_{xy}$ at a density of $n=2.37\times 10^{12} \text{ cm}^{-2}$ measured at $T=1.6$~K. The Hall resistivity is hysteretic, with a coercive field of several tens of millitesla, and we observe a well quantized $R_{xy}=h/e^2$ along with $R_{xx}<1 \mathrm{k\Omega}$ persisting through $B=0$ indicative of a QAH state stabilized by spontaneously broken time reversal symmetry.  In contrast to magnetically doped systems\cite{chang_experimental_2013,wang_quantum_2013,kou_scale-invariant_2014,checkelsky_trajectory_2014,chang_high-precision_2015,mogi_magnetic_2015,kou_mapping_2015}, the switching transitions are marked by discrete, $\Delta R\approx h/e^2$ steps in both $R_{xx}$ and $R_{xy}$, typical of magnetic systems consisting of a small number of domains\cite{yasuda_quantized_2017}. Figure 1C shows the detailed density evolution of the $R_{xy}$ hysteresis near $\nu=3$. Both the coercive field and zero-field Hall resistance are maximal near $\nu=3$, although hysteresis can be observed over a much broader range of $\nu\in(2.84,3.68)$ (see Fig.~\ref{fig:extentofhysteresis}).

\begin{figure*}[ht]
\includegraphics[width=4.75 in]{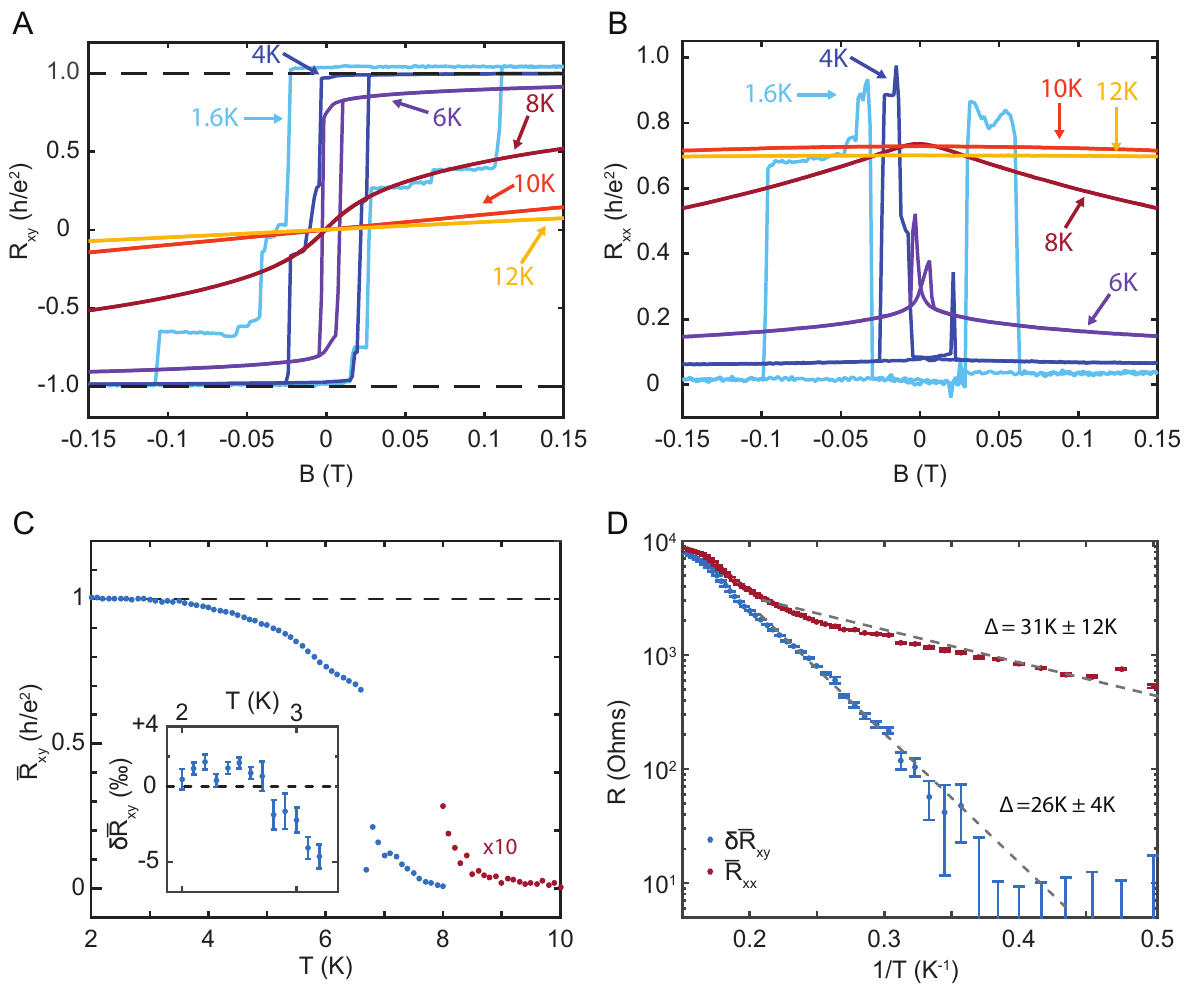}
\caption{\textbf{Temperature dependence of the quantum anomalous Hall effect.}
$(\textbf{A})$ $R_{xy}$  and $(\textbf{B})$ $R_{xx}$  as a function of $B$ measured at various temperatures for $n=2.37\times 10^{12} \text{ cm}^{-2}$. $R_{xx}$ and $R_{xy}$ mixing was corrected using contact symmetrization\cite{supplementary_information}.
$(\textbf{C})$ Temperature dependence of the field-training symmetrized resistance $\bar R_{xy}$ at $B=0$, described in the main text.  Data from $8$~K to $10$~K are multiplied by 10 so that the persistence of hysteresis up to the Curie temperature  $T_C\approx 9$~K is evident. The inset presents $\bar R_{xy}$ at $B=0$ at low temperatures, where it saturates for $T<2.7$~K to a value of $(1.001 \pm 0.0002)\times \frac{h}{e^2}$.  Error bars are the standard error derived from 11 consecutive measurements.
$(\textbf{D})$ Arrhenius plots of field training symmetrized resistances $\bar R_{xx}$ and $\delta \bar R_{xy}=h/e^2-\bar R_{xy}$. Dotted lines denote representative activation fits.  Systematic treatment of uncertainty arising from the absence of a single activated regime gives $\delta=31\pm 11$~K and $26\pm 4$~K for $\bar R_{xx}$ and $\delta \bar R_{xy}$, respectively\cite{supplementary_information}.
}
\label{fig:2}
\end{figure*}

Fig.~\ref{fig:1}D shows a schematic representation of the band structure at full filling ($\nu=4$) and at $\nu=3$.  In the absence of interaction-driven order, the spin-degenerate bands in each valley have total Chern number $\pm2$ (Fig.~\ref{fig:1}D).  The observed QAH state occurs because the exchange energy is minimized when an excess valley- and spin-polarized Chern band\cite{bultinck_anomalous_2019,zhang_twisted_2019} is occupied, spontaneously breaking time-reversal symmetry. Magnetic order in two dimensions requires anisotropy.  In graphene, the vanishingly small spin orbit coupling provides negligible anisotropy for the spin system.  It is thus likely that the  observed magnetism is orbital, with strong, easy-axis anisotropy arising from the two dimensional nature of the graphene bands\cite{xie_nature_2018,sharpe_emergent_2019,lu_superconductors_2019,bultinck_anomalous_2019,zhang_twisted_2019}.

The phenomenology of $\nu=3$ filling is nonuniversal across devices: some samples are metallic\cite{cao_correlated_2018,cao_unconventional_2018}, some\cite{yankowitz_electric_2014,lu_superconductors_2019} show a robust, thermally activated trivial insulator while others show an anomalous Hall effect\cite{sharpe_emergent_2019}. This is consistent with theoretical expectation\cite{xie_nature_2018} that the phase diagram at integer $\nu$ is highly sensitive to model details which, in our experiment, may be controlled by sample strain\cite{liu_nematic_2019} and alignment to an hBN encapsulant layer that breaks the $C_2$ rotation symmetry of tBLG\cite{bultinck_anomalous_2019,zhang_twisted_2019}.
The prior report of magnetic hysteresis at $\nu=3$ was indeed associated with close alignment of one of the two hBN encapsulant layers\cite{sharpe_emergent_2019}, a feature shared by our device\cite{supplementary_information}. Additional features of the transport phenomenology presented here further suggest that the single particle band structure of the device is significantly modified relative to unaligned tBLG devices, and suggest that hBN aligned samples constitute a different class of tBLG devices with distinct phenomenology.  First, our device shows only a weakly resistive feature at $\nu=2$, but a robust thermally activated insulator at charge neutrality. Remarkably, this $\nu=0$ insulator has a larger activation gap than even the states at $\nu=\pm4$, which are much smaller than typical\cite{supplementary_information}. Second, the quantum oscillations are highly anomalous, with hole-like quantum oscillations originating at $\nu=2$, again in contrast to all prior reports\cite{cao_correlated_2018,cao_unconventional_2018,yankowitz_tuning_2019,lu_superconductors_2019}.  While no detailed theory for these observations is available, the extreme sensitivity of the detailed structure of the flat bands to model parameters, combined with observations that hBN substrates can produce energy gaps as large as 30 meV in monolayer graphene\cite{hunt_massive_2013}, point to the role of the substrate in tipping the balance between competing many-body ground states  at $\nu=3$  in favor of the QAH state.

Figs. 2A and B show the temperature dependence of major hysteresis loops in $R_{xx}$ and $R_{xy}$, respectively.  As $T$ increases, we observe both a departure from resistance quantization and a suppression of hysteresis, with the Hall effect showing linear behavior in field by $T=12$~K.
In our measurments, we observe resistance offsets of $\sim 1$~$\mathrm{k\Omega}$ from the ideal value, which vanish when resistance is symmetrized or antisymmetrized with respect to magnetic field (or, for $B\approx0$, with respect to field training).  This is likely the result of a large contact resistance associated with one of the electrical contacts used\cite{supplementary_information}.
For quantitative analysis of the $T$-dependent data, we thus study field-training symmetrized resistances, denoted $\bar R_{xy}$ and $\bar R_{xx}$.  Figure 2C shows $\bar R_{xy}(0)$. We determine the Curie temperature to be $T_C\approx 9$~K from the onset of a finite $\bar R_{xy}(0)$, which indicates spontaneously broken time reversal symmetry.  At low temperatures, $\bar R_{xy}$ is quantized to $(1.001 \pm 0.0002) \times \frac{h}{e^2}$, remaining quantized up to $T=3$~K before detectable deviation is observed.

To quantitatively assess the energy scales associated with the QAH state, we measure the activation energy at low temperature.
Fig.~2D shows both the measured $\bar R_{xx}$ and the deviation from quantization of the Hall resistance, $\delta \bar R_{xy}=h/e^2-\bar R_{xy}$, on an Arrhenius plot.  We assume that the Hall conductivity $\sigma_{xy}$ is approximately $T$-independent and the longitudinal conductivity $\sigma_{xx}\sim e^{-\Delta/(2T)}$, where $\Delta$ is the energy cost of creating and separating a particle-antiparticle excitation of the QAH state.  Within this picture, inverting the conductivity tensor gives $\delta R_{xy}\sim e^{-\Delta/(T)}$ while  $R_{xx}\sim e^{-\Delta/(2T)}$\cite{supplementary_information}.
We find the activation gaps extracted from fitting $\delta \bar R_{xy}$ and $\bar R_{xx}$ to be $\Delta=26\pm 4$~K  and $\Delta=31\pm 12$~K, respectively, with the large uncertainty in the latter arising from the absence of a single simply activated regime\cite{supplementary_information}.
The activation energy is thus several times larger than $T_C$, in  contrast to magnetically doped topological insulator films where activation gaps are typically 10-50 times smaller than $T_C$\cite{supplementary_information, mogi_magnetic_2015, chang_high-precision_2015}.

\begin{figure}[ht]
\includegraphics[width=2.25 in]{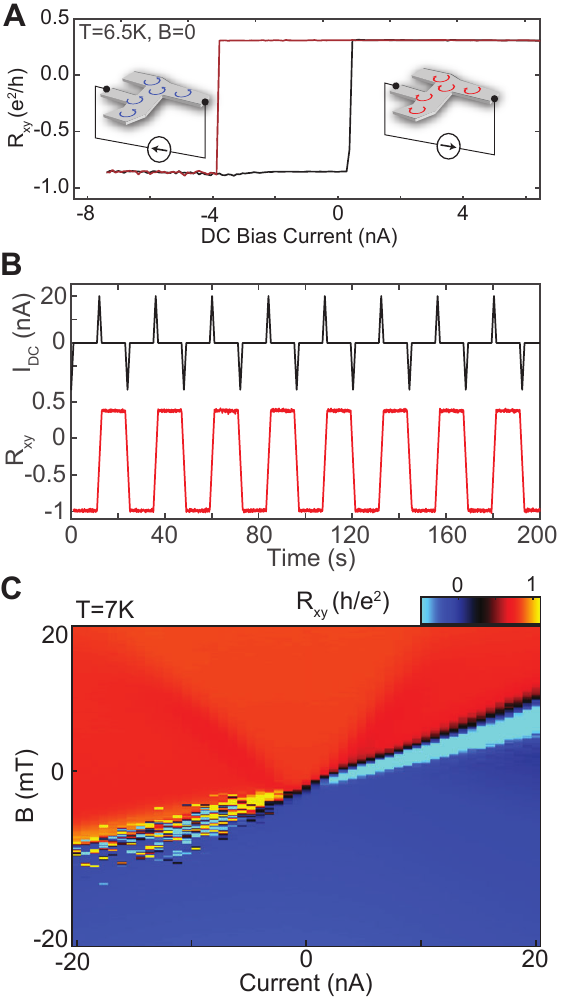}
\caption{\textbf{Current controlled magnetic switching.}
$(\textbf{A})$ $R_{xy}$ as a function of applied DC current, showing hysteresis as a function of DC current analogous to the response to an applied magnetic field.   Insets: schematic illustrations of current controlled orbital magnetism.
$(\textbf{B})$ Nonvolatile electrical writing and reading of a magnetic bit at $T=6.5$~K and $B=0$.  A succession of 20 nA current pulses of alternating signs controllably reverses the magnetization, which is read out using the AC Hall voltage. The magnetization state of the bit is stable for at least $10^3$ s \cite{supplementary_information}.  $(\textbf{C})$ $R_{xy}$ as a function of both DC bias current and magnetic field at $7$~K.  Opposite directions of DC current preferentially stabilize opposite magnetization states of the bit. Measurements presented in (\textbf{A} - \textbf{C}) are neither field nor Onsager symmetrized, which produces an offset in $R_{xy}$.}
\label{fig:3}
\end{figure}

Ferromagnetic domains in tBLG interact strongly with applied current\cite{sharpe_emergent_2019}.  In our device, this allows deterministic electrical control over domain polarization using exceptionally small DC currents.  Figure \ref{fig:3}A shows $R_{xy}$  at $6.5$~K and B=0, measured using a small AC excitation of $\sim100$~pA to which we add a variable DC current bias.  We find that the applied DC currents drive switching analogous to that observed in an applied magnetic field, producing hysteretic switching between magnetization states.  DC currents of a few nanoamps are sufficient to completely reverse the magnetization, which is then indefinitely stable\cite{supplementary_information}.  Figure \ref{fig:3}B shows deterministic writing of a magnetic bit using current pulses, and its nonvolatile readout using the large resulting change in the anomalous Hall resistance. High fidelity writing is accomplished with 20~nA current pulses while readout requires $<100$~pA of applied AC current.

Assuming a uniform current density in our micron-sized, two atom thick tBLG device results in an estimated current density $J<10^4$~A$\cdot$cm$^{-2}$.  While current-induced switching at smaller DC current densities has been realized in MnSi, readout of the magnetization state in this material has so far only been demonstrated using neutron scattering\cite{jonietz_spin_2010}.
Compared with other systems that allow in situ electrical readout, such as GaMnAs\cite{jiang_efficient_2019} and Cr-(BiSb)$_2$Te$_3$ heterostructures\cite{fan_magnetization_2014}, the applied current densities are at least one order of magnitude lower. More relevant to device applications, the absolute magnitude of the current required to switch the magnetization state of the system  ($\sim 10^{-9}$A) in our devices is, to our knowledge, 3 orders of magnitude smaller than reported in any system.

Figure \ref{fig:3}C shows the Hall resistance at $T=7$~K measured as a function of magnetic field and current.  Applied DC currents apparently compete directly with the applied magnetic field: as shown in the figure, opposite signs of current stabilize opposite magnetic polarizations, and can stabilize states aligned opposite to that favored by the applied field. We note that, while current does break time reversal symmetry, the observed behavior is not compatible with mirror symmetry across the plane perpendicular to the sample and parallel to the net current flow, assuming the injected current is not spin- or valley-polarized.
We propose instead a simple mechanism for the low-current switching that arises from the interplay of edge state physics and device asymmetry\cite{supplementary_information}.  In a QAH state, an applied current generates a chemical potential difference between the chiral one dimensional modes located on opposite sample edges.  Due to the opposite dispersion of a given edge state in opposite magnetic states (which have opposite $C$), the DC current $I$ raises (or lowers) the energy of the system by $\delta E\sim \pm \frac{8\pi^2}{3} \frac{\hbar^2}{m e^3 v^3}L I^3$
for a $C=\pm1$ state, where $m$ and $v$ are the edge state effective mass and velocity, $e$ is the elementary charge, and $L$ is the length of the edge state.
When the edges have different lengths or velocities, the current favors one of the two domains, with the sign and magnitude of the effect dictated by the device asymmetry. For a current in the range of $I=10-100$~nA, comparable to the switching currents observed at low temperatures, using estimates of $m$ and $v$ based on bulk measurements\cite{supplementary_information} and assuming an edge length difference of $\approx 1$~$\mathrm{\mu m}$. gives $\delta E$ comparable to the magnetic dipole energy due to a 1mT field.

We note that while this effect should be  generic to all QAH systems, it is likely to be dominant at low currents in tBLG due to the weak pinning of magnetic domains and small device dimensions.  Crucially, it provides an engineering parameter for electrical control of domain structure that can be deterministically encoded in the device geometry, enabling new classes of magnetoelectric devices.

\let\oldaddcontentsline\addcontentsline% Store \addcontentsline
\renewcommand{\addcontentsline}[3]{}% Make \addcontentsline a no-op
\begin{acknowledgments}
\textbf{Acknowledgements:} The authors acknowledge discussions with A. Macdonald, Y. Saito, and M. Zaletel.
Device fabrication was supported by the ARO under W911NF-17-1-0323.
Measurements were supported by the AFOSR under FA9550-16-1-0252.
CT acknowledges support from the Hertz Foundation and from the NSF GRFP under Grant No. 1650114.
L.B. was supported by the  DOE  Office  of  Sciences  Basic  Energy  Sciences program under Award No. DE-FG02-08ER46524.
AFY acknowledges the support of the David and Lucille Packard Foundation and the Alfred P. Sloan foundation.

\end{acknowledgments}

\let\addcontentsline\oldaddcontentsline% Restore \addcontentsline

% \section{Author Contributions: }

\let\oldaddcontentsline\addcontentsline% Store \addcontentsline
\renewcommand{\addcontentsline}[3]{}% Make \addcontentsline a no-op
\bibliographystyle{unsrt}

%\bibliography{references,NewReferences}
%\bibliography
\let\addcontentsline\oldaddcontentsline% Restore \addcontentsline

\clearpage

%%%%%%%%%%%%%%%%%%%%%%%%%%%%%%%%%%%%%%%%%%%%%%%%%%%%%%%%%%%%%%%%%%%%%%%%%%%%%%%%%%%%%%%
\pagebreak
\widetext
\begin{center}
\textbf{\large Supplementary Material for: Intrinsic quantized anomalous Hall effect in a moir\'e heterostructure}
\end{center}
\renewcommand{\thefigure}{S\arabic{figure}}
\renewcommand{\thesubsection}{S\arabic{subsection}}
\setcounter{secnumdepth}{2}
\renewcommand{\theequation}{S\arabic{equation}}
\renewcommand{\thetable}{S\arabic{table}}
\setcounter{figure}{0}
\setcounter{equation}{0}
\onecolumngrid
\appendix
\tableofcontents
\section{Experimental Methods}

\subsection{Device fabrication}

\begin{figure*}[h]
\includegraphics[width=4.75 in]{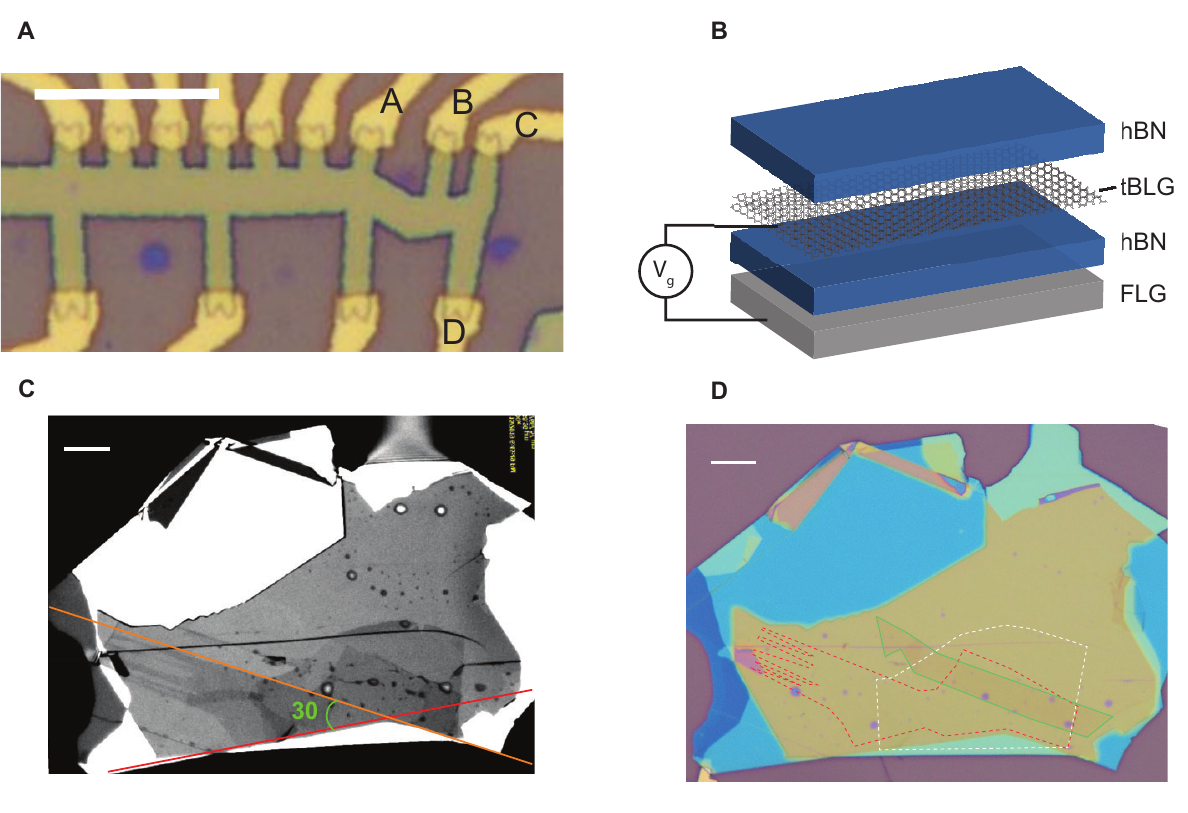}
\caption{
\textbf{tBLG device.}
(\textbf{A}) Optical micrograph of the device. Scale bar corresponds to 10 $\mu m$.
(\textbf{B}) Schematic of a tBLG heterostructure. tBLG is encapsulated between flakes of hBN, with a flake of few-layer graphite used as a gate.
(\textbf{C}) Optical image of the stack. The crystalline edge of the top hBN and the top layer of tBLG are aligned with a 30 degree angular offset in the marked cleavage planes.
(\textbf{D}) Optical image of the stack before etching, showing the top layer of the tBLG (red dashed line), bottom layer of the tBLG (white dashed line), and the bottom gate (green solid line).
}
\label{fig:OpticalImage}
\end{figure*}

An optical micrograph of the tBLG device discussed in the main text is shown in Fig. \ref{fig:OpticalImage}A . The device is made using the ``tear-and-stack'' technique~\cite{kim_van_2016}. The tBLG layer is sandwiched between two  hBN flakes with thickness 40 and 70~nm, as shown in Fig. \ref{fig:OpticalImage}B. A few-layer-thick graphite flake is used as the bottom gate of the device, which has been shown to produce devices with low charge disorder~\cite{zibrov_tunable_2017}. The stack rests on a Si/SiO$_2$ wafer, which is also used to gate the contact regions of the device. The stack was assembled at 60$^\circ$~C using a dry-transfer  technique~\cite{wang_one-dimensional_2013} with a poly(bisphenol A carbonate) (PC) film on top of a polydimethylsiloxane (PDMS) stamp.

To determine the twist angle of the device, we first find the conversion factor from applied gate voltage to density by measuring the period of Shubnikov de Haas oscillations in Fig.\ref{fig:LandauFanLowT}. We then identify the densities at which, at $B=0$, features appear we presume to be associated with commensurate band filling.  From measurements of $n$ and $\nu$, we then determine the angle by the relation $n/\nu=1/A_m \simeq 2 \theta^2 / \sqrt{3}a^2$, where $a$= 0.246 nm is the lattice constant of graphene. Using the clear resistivity peak at $\nu=0$ and $2$, the $R_{xy}$ plateau at $\nu=3$, and the edge of the insulating regions at $\nu=\pm4$ gives $\theta=1.15\pm.01^\circ$.

In an exfoliated heterostructure, the orientation of the crystal lattice relative to the edges of the flake can often be determined by investigating the natural cleavage planes of the flake. Honeycomb lattices, like graphene and hBN, have two natural cleavage planes, each with six-fold symmetry, that together produce cleaveage planes for every 30 degree relative rotation of the lattice. In an optical image of the device, we identify crystalline edges on both the graphene and the top hBN flakes by finding edges with relative angles of 30 degrees. Using this method, the tBLG layer in the device was visually aligned with the top hBN as shown in Fig.~\ref{fig:OpticalImage}C. This procedure produces aligned hBN and graphene crystalline lattices only if the same natural cleavage plane was chosen for both the hBN and graphene flake. It is impossible to know without more sophisticated methods exactly which of the two cleavage planes was selected. However, when aligned, the hBN forms an additional moir\'e pattern with the tBLG, which in turn produces an additional miniband that exhibits resistive peaks when fully filled or emptied.  These features can be observed in transport and, when they occur, can confirm that a graphene flake has been successfully aligned to an adjacent BN. We observe both a small additional resistive peak in the longitudinal resistance at $n = 2.7 \times 10^{12} \text{ cm}^{-2}$ (Fig. \ref{fig:1}) and an additional feature in the field dependent quantum oscillations emanating from $\nu=4$ in Fig.~\ref{fig:LandauFan}. Both observations are consistent with either hBN-graphene alignment of approximately $0.4^\circ$\cite{hunt_massive_2013} or a tBLG region with twist angle $\theta \approx 1.1^\circ$. We do not observe equivalent features at opposite density, which would be expected in either case. The only other observation to date of zero field ferromagnetism in a tBLG sample reported both crystallographic alignment and additional insulating peaks, symmetric in density, associated with the hBN-graphene moir\'e lattice\cite{sharpe_emergent_2019}.

\subsection{Electrical transport measurements}

Four-terminal resistance measurements are carried out in a liquid helium cryostat with a $1$~K pot and a base temperature of $1.6$~K. The measurement is done using AC current excitations of 0.1 - 20 nA at 0.5 - 5.55 Hz using a DL 1211 current preamplifier, SR560 voltage preamplifier, and SR830 and SR860 lock-in amplifiers.  Gate voltages and DC currents are applied, and amplified voltages recorded, with a home built data acquisition system based on AD5760 and AD7734 chips.
The contacts of the device that show a robust quantized anomalous Hall signal are labeled A, B, C, and D in Fig.~\ref{fig:OpticalImage}A. In this section, we use the following convention for labeling contact configurations for resistance measurements: $R_{ABCD}$ corresponds to a resistance measured by applying current from A to B and measuring the voltage between C and D.

\subsubsection*{Disentangling longitudinal and Hall resistance}

The geometry of the device is such that no pair of contacts allows a perfectly isolated determination of $R_{xx}$ or $R_{xy}$, and measurements must be performed in a van der Pauw method. Magnetic field and geometric symmetrization techniques were thus used to separate these values with quantitative precision\cite{sample_reversefield_1987}.

Field symmetrization relies on the fact that longitudinal resistances are symmetric with respect to time reversal, whereas Hall resistances are antisymmetric. Above the coercive field, pairs of resistance measurements performed at opposite fields can thus be used to extract the longitudinal and Hall resistances using:
\begin{align}
    R_{xx}(B) = (R_{\text{meas}}(B) + R_{\text{meas}}(-B) )/2 \\
    R_{xy}(B) = (R_{\text{meas}}(B) - R_{\text{meas}}(-B) )/2
\end{align}

\begin{figure*}[h]
\includegraphics[width=4.75 in]{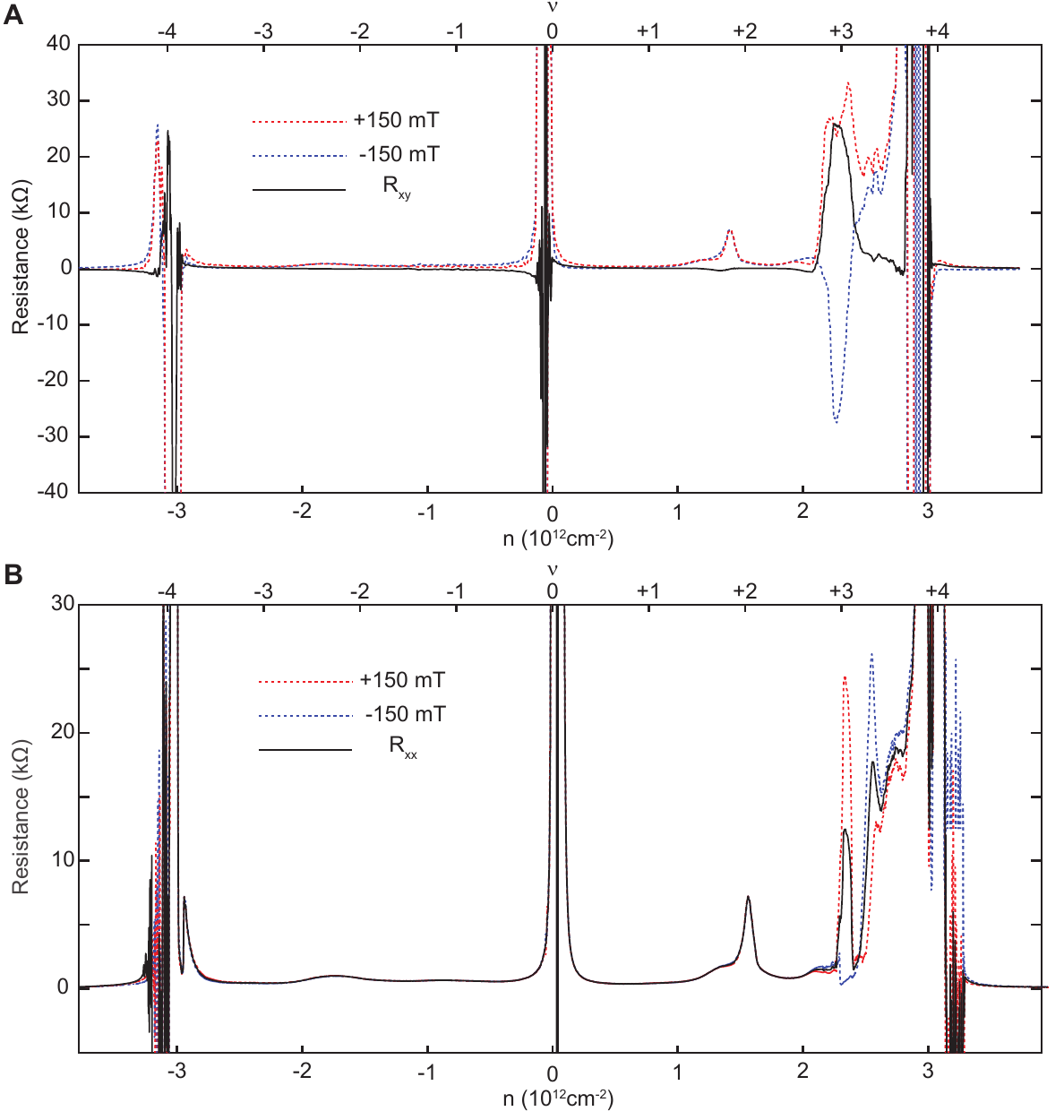}
\caption{\textbf{Measurements of $R_{xx}$ and $R_{xy}$ using reverse field reciprocity.}
(\textbf{A})  $R_{ACBD}$ measured at +150 mT and -150 mT. $R_{xy}$ is obtained from the antisymmetric part of the resistance.
(\textbf{B})  $R_{ADBC}$ measured at +150 mT and -150 mT. $R_{xx}$ is obtained from the symmetric part of the resistance.
}
\label{fig:fieldsymillust}
\end{figure*}

In Fig. \ref{fig:fieldsymillust}, we show an example of this symmetrization method. $R_{xx}$ and $R_{xy}$ were acquired by extracting the field symmetric and field antisymmetric parts of $R_{ADBC}$ and $R_{ACBD}$ at $B = \pm$150~mT. When time reversal symmetry is broken independently of the applied magnetic field---as in the case of the observed ferromagnetism---an analogous symmetrization technique can be used by symmetrizing and antisymmetrizing the measured resistance at $B = 0$ for opposite signs of magnetic field training .  This method has the advantage of not requiring changes to the measurement contacts, which in practice allows for more precise determination of the components of the resistivity tensor. However, this form of symmetrization does not accurately capture domain dynamics of the device that are asymmetric in field.

For faithful measurements of the longitudinal and Hall resistances as a function of $B$, we instead use Onsager symmetrization. In an isotropic sample, Hall resistances appear in the resistance tensor as off-diagonal antisymmetric terms:

$$ R =
\quad
\begin{pmatrix}
R_{xx} & R_{xy} \\
-R_{xy} & R_{xx}
\end{pmatrix}
\quad
$$

By exchanging the voltage contacts with the current contacts we can measure the transpose of the resistance tensor, and can therefore symmetrize by using the following formulas, which remain valid for all systems in the linear response regime\cite{sample_reversefield_1987} :

\begin{align}
    R_{xx} = (R_{\text{ABCD}} + R_{\text{CDAB}})/2 \\
    R_{xy}(B) = (R_{\text{ABCD}} - R_{\text{CDAB}}) )/2
\end{align}

\begin{figure*}[ht]
\includegraphics[width=7.25 in]{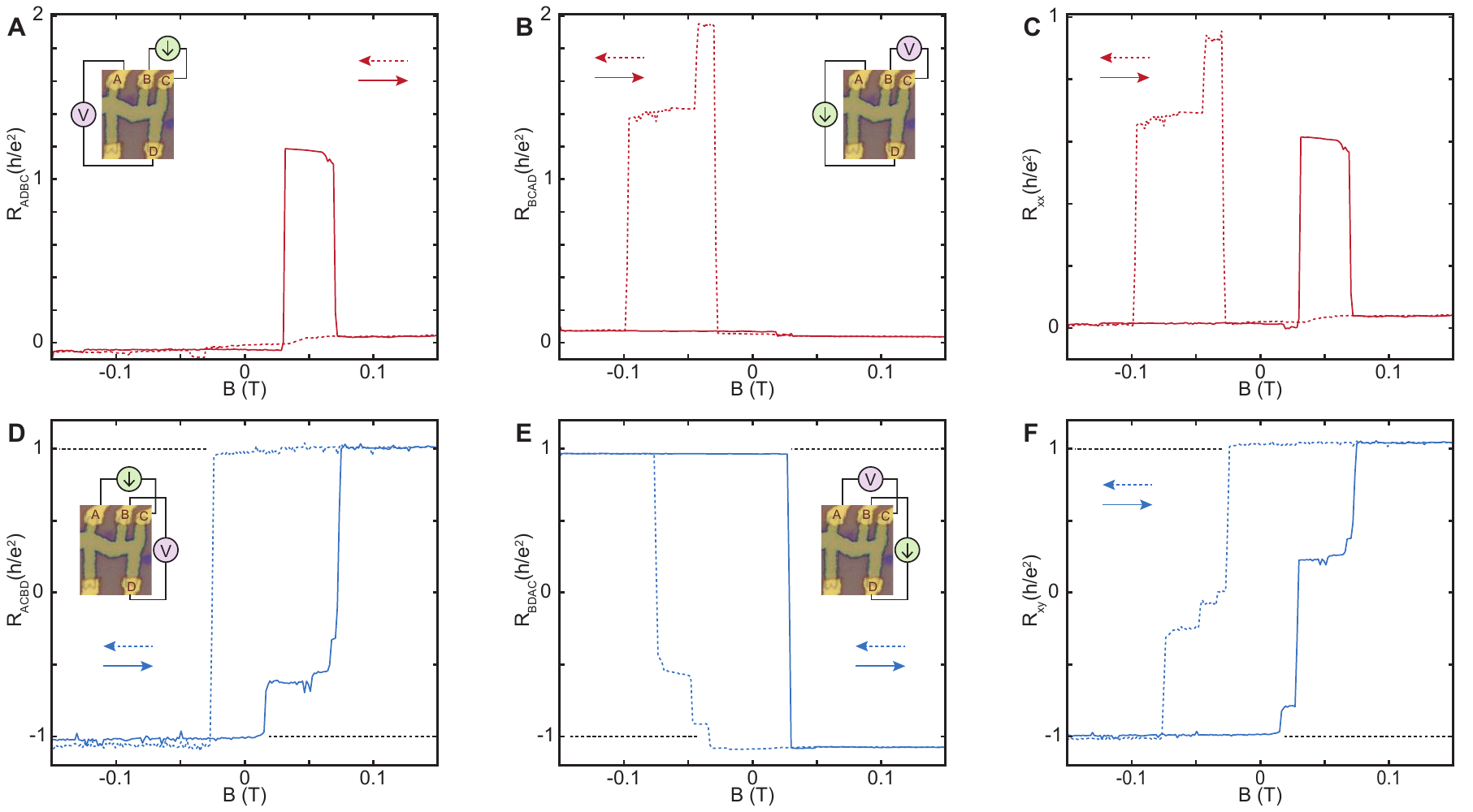}
\caption{\textbf{Measurements of $R_{xx}$ and $R_{xy}$ as functions of magnetic field using Onsager reciprocal relation.} (\textbf{A}) Resistance measurement applying current from contact A to D and measuring voltage from contact B to C. The resistance peak is only observed in the positive field sweep direction. (\textbf{B}) Resistance measurement applying current from contact B to C and measuring voltage from contact A to D. The resistance peak is only observed in the negative field sweep direction. The peak value of the longitudinal resistance is close to $2h/e^2$.  Both of these phenomena are signatures of two QAH domains in series~\cite{yasuda_quantized_2017}. (\textbf{C}) Symmetrized $R_{xx}$ obtained from (\textbf{A}) and (\textbf{B}) using $R_{xx} = 1/2 \times (R_{ADBC} + R_{BCAD})$. (\textbf{D}) Resistance measurement applying current from contact A to C and measuring voltage from contact D to B. (\textbf{E}) Resistance measurement applying current from contact D to B and measuring voltage from contact A to C. (\textbf{F}) Symmetrized $R_{xx}$ obtained from (\textbf{D}) and (\textbf{E}) using $R_{xy} = 1/2 \times (R_{ACBD} - R_{BDAC})$. Opposite sweep directions are marked with solid and dashed lines.
}
\label{fig:Onsager}
\end{figure*}

In practice, we expect different contact configurations to carry larger longitudinal or Hall resistance signals based on the geometry of the sample. $R_{\text{ADBC}}$ is predominantly a measurement of $R_{xx}$, whereas $R_{\text{ACBD}}$ is the contact configuration that maximizes the relative contribution of $R_{xy}$. In Fig. \ref{fig:Onsager} A-C, we use $R_{ADBC}$ and its Onsager reciprocal $R_{\text{BCAD}}$ to extract $R_{xx}$, and in Fig. \ref{fig:Onsager} D-F we use $R_{\text{ACBD}}$ and its Onsager reciprocal $R_{\text{BDAC}}$ to extract $R_{xy}$. This measurement modality has the advantage of providing full field dependence of longitudinal and Hall resistances.  However, because it entails switching physical contact connections, symmetrization by this method can introduce sensitivity to offsets associated with high contact resistances on certain contacts, discussed in the next section.

We notice that for each configuration used for $R_{xx}$ measurements, the resistance peak associated with a domain transition is only observed in one sweep direction. At low temperatures, we observe the unsymmetrized $R_{xx}$ peak approach $2h/e^2$ during the transition. Both facts are consistent with the magnetic reversal occuring through an intermediate state in which two magnetic domains with opposite polarization are arranged in series along the direction of current flow \cite{yasuda_quantized_2017}.

\vspace{-15pt}
\subsubsection*{Offsets and Contact Resistances}

\begin{figure*}[ht]
\includegraphics[width=7.25 in]{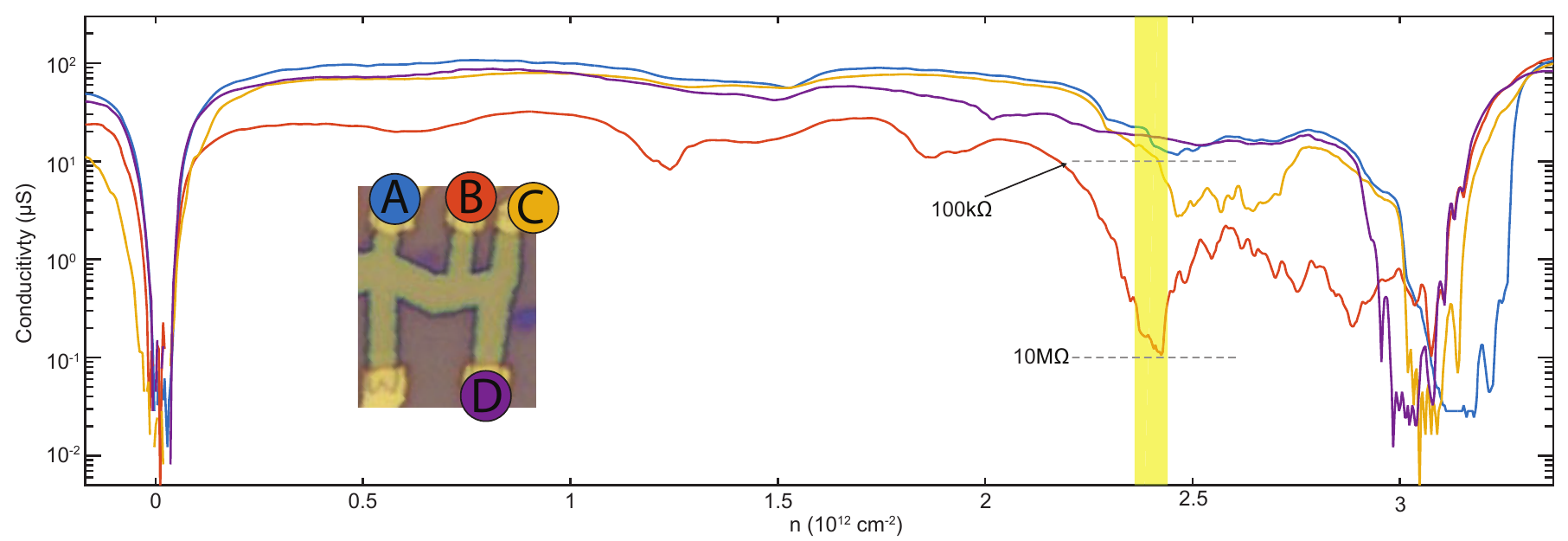}
\caption{\textbf{Characterization of device electrical contacts.} Density dependence of the contact conductivity of the four contacts used in the experiment at $1.6$~K. We perform a two terminal measurement sourcing current from one contact with all of the remaining contacts grounded. The density region at which the QAH effect is observed is highlighted in yellow. Contact B is two orders of magnitude more resistive than all other contacts close to $\nu=3$.}
\label{fig:Contacts}
\end{figure*}

We attribute small offsets in longitudinal and Hall resistances to a highly resistive electrical contact to the device. In Fig. \ref{fig:Contacts}, we show the density dependence of the two-terminal conductivity of the contacts used in the measurement with current sourced from one contact and drained through the rest. In all contact configurations we have the expected insulating states at charge neutrality and full filling of the miniband, with resistive features near $\nu = 2$ and $3$. However, in contact B we observe another strongly insulating state near $\nu=3$, at the densities at which we observe the QAH effect. We attribute this to a tBLG region in series with contact B that exhibits a strongly insulating $C=0$ phase--potentially arising due to misalignment or weak coupling to the proximal hBN layers. When this contact is used as a current source, near $\nu=3$ its series resistance is high enough to introduce offsets in $R_{xy}$.  When this contact is used as a voltage probe, its weak coupling to the rest of the device makes the resulting measurement of $R_{xy}$ noisy.  The measurement scheme that was used to quantitatively assess the quantization of the QAH state in Fig. \ref{fig:2} was a field symmetrized measurement of $R_{xy}$ using contact B as a voltage probe and very long averaging times.

\vspace{-15pt}
\subsubsection*{Quantum anomalous Hall thermal activation measurements} \label{sec:test}
We obtain the thermal activation gap of $\sigma_{xx}$ by fitting both the longitudinal resistance, $R_{xx}$, and the deviation from quantization in the Hall resistance, $\delta \bar R_{xy}$. We model $\sigma_{xx}$ as thermally activated with $\sigma_{xy}$ remaining constant. In the limit of small $\sigma_{xx}$, this gives:
\begin{align}
    \rho_{xx} = \sigma_{xx} / \left(\sigma_{xx}^2 + \sigma_{xy}^2 \right) \approx \sigma_{xx} / \sigma_{xy}^2\\
    \rho_{xy} = \sigma_{xy} / \left(\sigma_{xx}^2 + \sigma_{xy}^2 \right) \approx 1 / \sigma_{xy} - \sigma_{xx}^2 / \sigma_{xy}^4
\end{align}

Assuming that the Fermi level lies in the middle of the gap, we take the gap size to be twice the measured thermal activation energy scale. We therefore fit the following equations to measured values of $R_{xx}(T)$ and $R_{xy}(T)$ to extract the gap size:
\begin{align}
\sigma_{xx} \propto R_{xx} = R_0 e^{- (\frac{\Delta}{2T})} \\
\sigma_{xx}^2 \propto h/e^2 - R_{xy} = R_1 e^{- (\frac{\Delta}{T})}
\end{align}

\begin{figure*}[ht]
\includegraphics[width=4.75 in]{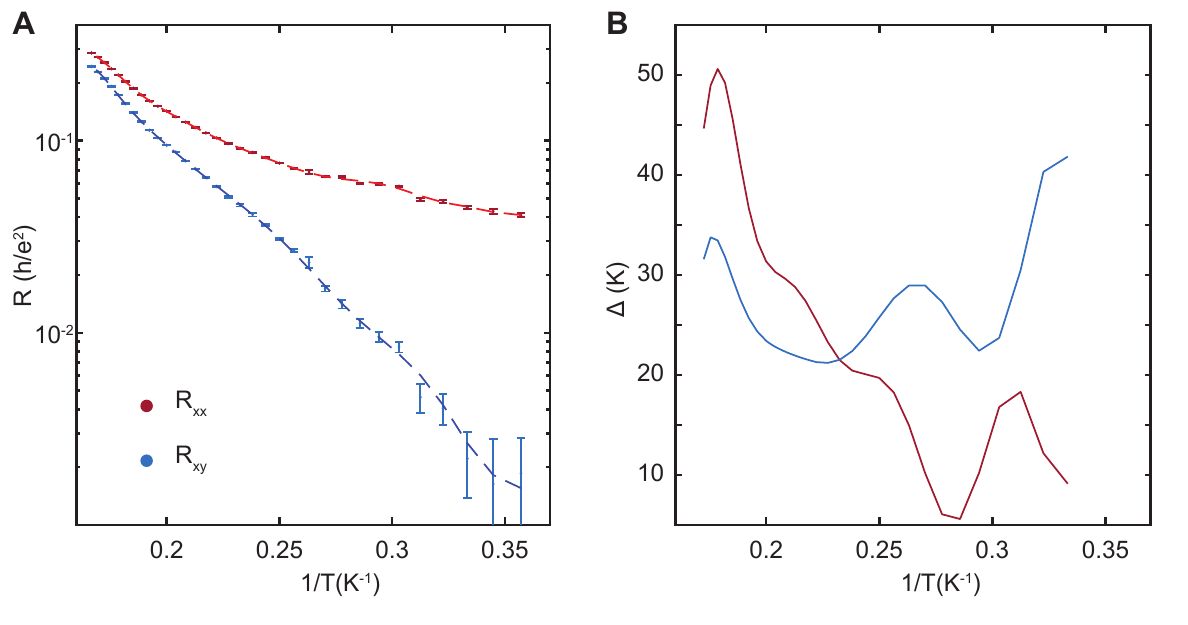}
\caption{\textbf{Thermal activation fit}
(\textbf{A}) Polynomial fit of  $\log \bar R_{xx}(\beta)$ and $\log(\delta \bar R_{xy}(\beta)$, where $\beta=1/T$. $\bar R_{xx}$ is fit with a thirteenth-degree polynomial and $\chi^2$ is 2.0. $\delta \bar R_{xy}$ is fit with a tenth-degree polynomial and $\chi^2$ is 1.8.
(\textbf{B}) Gap size extracted from derivative of polynomial fit in (\textbf{A}). Gap over the temperature range is taken as the weighted average with weight proportional to the log scale error at each temperature point. The first and last two temperature points are removed because the local derivative cannot be well approximated without symmetric sampling around the temperature of interest. We obtained a gap value of $31 \text{ K} \pm 12$~K from the $R_{xx}$ data and of $26 \text{ K} \pm 4$~K from the $\delta \bar R_{xy}$ data.
}
\label{fig:GapError}
\end{figure*}

Arrhenius plots of $R_{xx}(T)$ and $h/e^2 - R_{xy}(T)$ have fluctuating slopes; as in most thermal activation measurements, the dominant uncertainty in the gap measurement arises from uncertainty in which temperature regime to fit.
For the gaps presented in Figure 2D in the main text, we fit a polynomial of order $n$ to log$(R)$, where $n$ is chosen such that the reduced $\chi^2$ goodness of fit parameter $\chi^2 - 1$ is minimized. We then take the weighted average and standard deviation of the derivative at each temperature point, with weight inversely proportional to the log scale error of the measurement. Fits are shown in Fig. \ref{fig:GapError} and are labeled with the gap obtained from the fit.

\newpage \clearpage

\section{Supplementary Data}

\subsection{Evidence for unusual band structure}

Theories predicting the emergence of the QAH effect in twisted bilayer graphene require either explicitly broken two-fold rotation symmetry, or the inclusion of remote conduction and valence bands\cite{bultinck_anomalous_2019, zhang_twisted_2019, xie_nature_2018}. The former, in particular, can be achieving by aligning the twisted bilayer graphene with the hBN substrate, which is known in monolayer graphene to generate large energy gaps at $B=0$\cite{hunt_massive_2013,amet_insulating_2013}.  In this section we present evidence that the sample in our study is likely aligned with the hBN. In particular, it shows an uncharacteristic hierarchy of integer $\nu$ gaps, with the largest observed gap at $\nu=0$.  The Landau fan measurement $R_{xx}(B,n)$ also shows unusual features that suggest a significantly modified electronic structure within the flat bands and possible signs of a second moire pattern.

\subsubsection*{Activation gaps at $\nu=0$ and $\nu=\pm4$}

\begin{figure*}[ht]
\includegraphics[width=2.25 in]{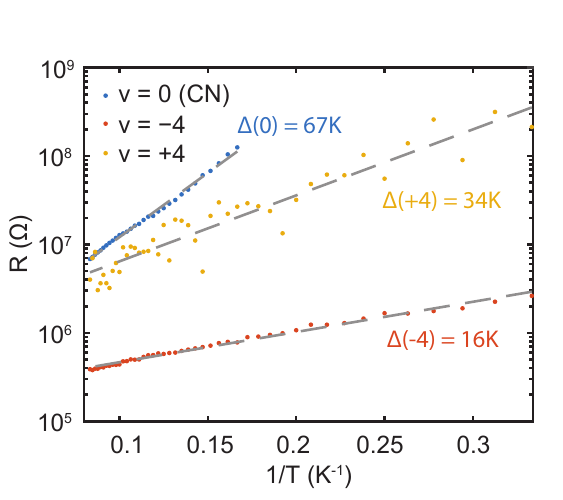}
\caption{\textbf{Insulating state activation gaps.}
Arrhenius plots presenting the temperature dependence of the resistivity of the insulating states at charge neutrality ($\nu = 0$) and $\nu=\pm 4$. Fits of the form $R \propto \mathrm{exp}(\Delta / 2k_BT)$ are used to extract the thermal activation gaps.}
\label{fig:Insulating}
\end{figure*}

In Fig. \ref{fig:Insulating}, we present thermal activation gap measurements at $\nu = 0$ and $\nu = \pm4$. The gap at charge neutrality is determined to be 67~K, much larger than other devices, though not unexpectedly so. However, the gaps at $\nu = \pm 4$ are uncharacteristically small with values of $34$~K for $\nu = 4$ and $16$~K for $\nu = -4$. The asymmetry in gap size is not seen in other devices\cite{polshyn_phonon_2019}.

\subsubsection*{Quantum oscillations}
Sufficiently low disorder 2D electron gases exhibit the quantum Hall effect at high magnetic fields, wherein the band structure is reshaped into Landau levels, topological bands that are gapped by the cyclotron energy. At intermediate magnetic fields, oscillations corresponding to incipient but not fully gapped Landau level plateaus appear as minima in $R_{xx}$ as a function of magnetic field and density. States with nontrivial topology, either from Landau level formation or instrinsic band topology, evolve with magnetic field according to the Streda formula:

\begin{align}
C = \frac{h}{e} \frac{dn}{dB}
\end{align}

\begin{figure*}[ht]
\includegraphics[width=7.25 in]{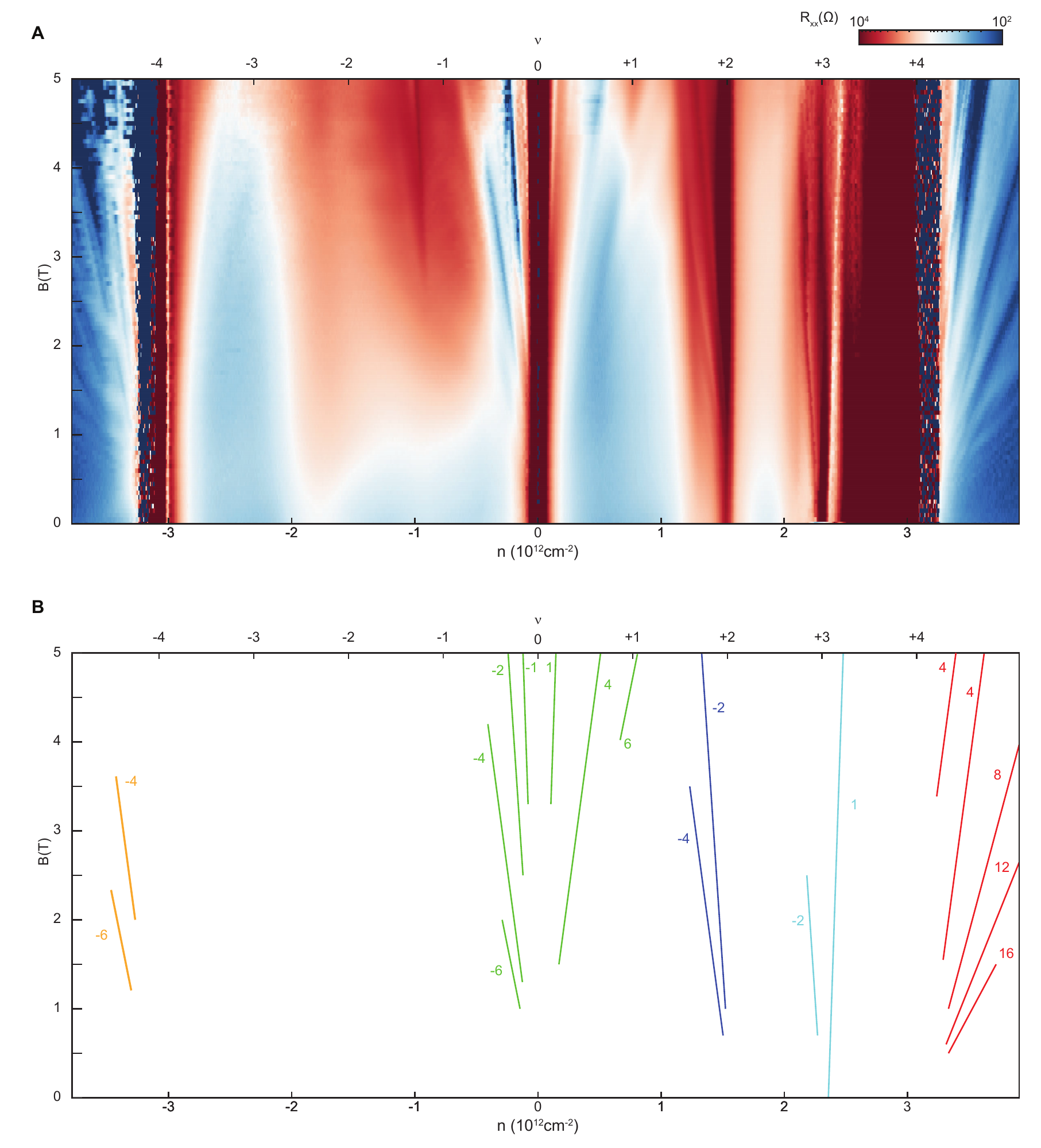}
\caption{\textbf{Quantum oscillations in QAH region.}
(\textbf{A}) Magnetic field and density dependence of $R_{xx}$ in the device at $1.6$~K. Data are taken from -5T to 5T and symmetrized with respect to magnetic field. (\textbf{B}) Schematic lines representing the evolution of minima in $R_{xx}(n,B)$ associated with topologically nontrivial bands.  Green lines labelling $C$ = $\pm$1, $-$2, $\pm$4, $\pm$6 mark quantum oscillations around the charge neutrality point. Similar sketches mark quantum oscillations around $\nu = -4$ (orange, $C$ = -4, -6), $\nu = 2$ (dark blue, $C$ = -2, -4), $\nu = 3$ (light blue, $C$ = -2, 1), and $\nu = 4$ (red, $C$ = 4, 4, 8, 12, 16). The two $C$ = 4 states marked in red near $\nu = 4$ are inconsistent with a single moir\'e pattern, implying either hBN-graphene alignment or disorder in the tBLG twist angle. The $C$ = 1 state marked in light blue corresponds to the QAH state.
}
\label{fig:LandauFan}
\end{figure*}

\begin{figure*}[ht]
\includegraphics[]{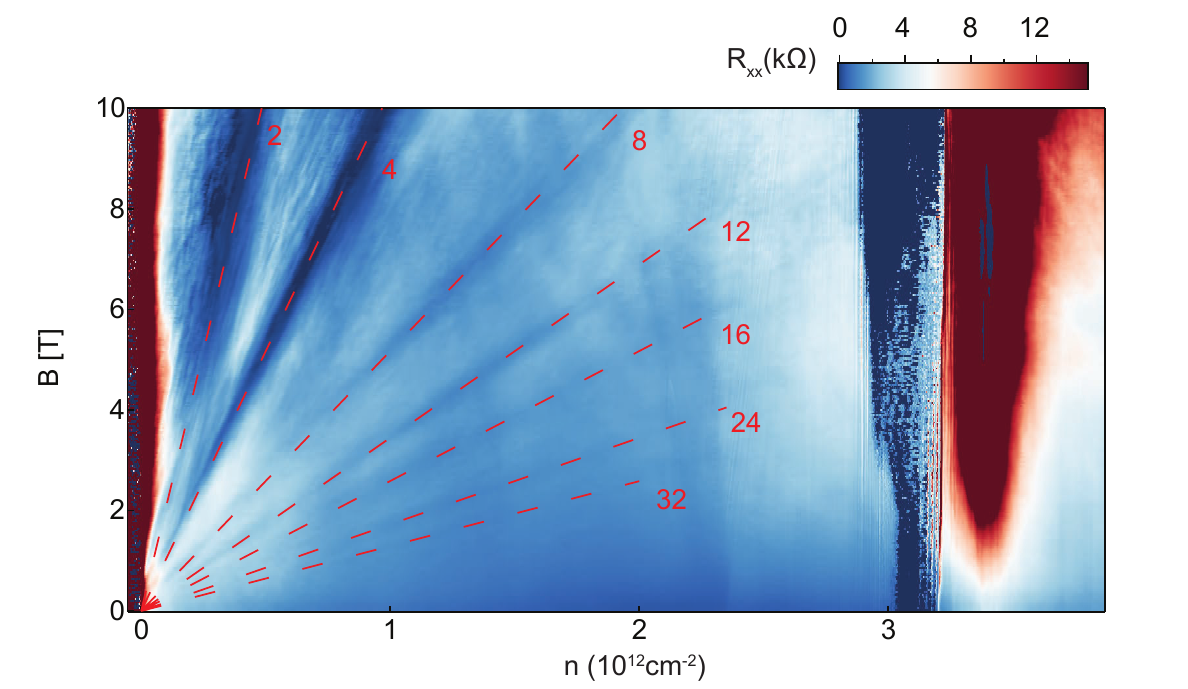}
\caption{\textbf{Quantum oscillations measured at low temperature.}
Quantum oscillations measured between contacts A and B at $30$~mK. The QAH effect is not visibile in this region of the device. Schematic lines representing the evolution of Landau Levels used to extract the gate capacitance are marked with dashed red lines.
}
\label{fig:LandauFanLowT}
\end{figure*}

Figures \ref{fig:LandauFan}A and \ref{fig:LandauFanLowT} show $R_{xx}$ as a function of density and magnetic field for this tBLG device. By measuring the slope of the resistance minima as a function of density and magnetic field, we obtain the Chern number $C$ of the set of filled bands at that density. This characterizes the topology of both the emergent topological bands (the Landau levels) and the intrinsically topological band (the QAH state).

By measuring $\frac{dn}{dB}$ for a set of Landau levels with ubiquitous, well-characterized behavior, we can determine the density within the device as a function of applied gate voltage. Figure \ref{fig:LandauFanLowT} shows quantum oscillations near charge neutrality measured at $30$~mK between contacts A and B, a region not exhibiting the QAH effect. The Landau Fan diagram shows clear gapped states with fillings $\nu=$ +2, 4, 8, 12, 16, 24, 32. We use these quantum oscillations to determine the capacitance of the device $C=4.1\times 10^{-4}$~$\mathrm{F/m^2}$. This extracted capacitance was used to populate the density axis wherever required.

Figure \ref{fig:LandauFan} shows quantum oscillations measured at $1.6$~K between contacts B and C. The QAH state can be seen at $n \sim 2.4 \times 10^{12}\text{ cm}^{-2}$, $B = 0$.  We can extract its Chern number from the Streda formula applied to its evolution as a function of magnetic field, and we find that $C = +1$ (\ref{fig:LandauFan}B).  This is in reassuring agreement with the measurement of $R_{xy}$ in transport measurements of the QAH state, for which $R_{xy} = \frac{1}{C} \frac{h}{e^2} = \frac{h}{e^2}$ in this device.  Quantum oscillations originating from the resistive state at $\nu = 2$ have hole-like character ($C = -2, -4, -6$), which has not been observed in other flat band tBLG devices thus far.

The quantum oscillations also show additional features that do not appear consistent with a single moir\'e pattern, particularly a weak fan in the second subband emanating from near $\nu=4$.  These features may arise from either inhomogeneity in the twist angle between the two graphene layers, or from a closely aligned proximal hBN layer.  Assuming the latter, the extracted moir\'e wavelength of 12.7 nm corresponds to a graphene/hBN misalignment of 0.46$^\circ$, while assuming the former implies a region of the twisted bilayer with 1.11$^\circ$ interlayer twist angle.

\clearpage

\subsection{Further characterization of hysteretic response }

\subsubsection*{Magnetic hysteresis as a function of density}

\begin{figure*}[h!]
\includegraphics[width=6 in]{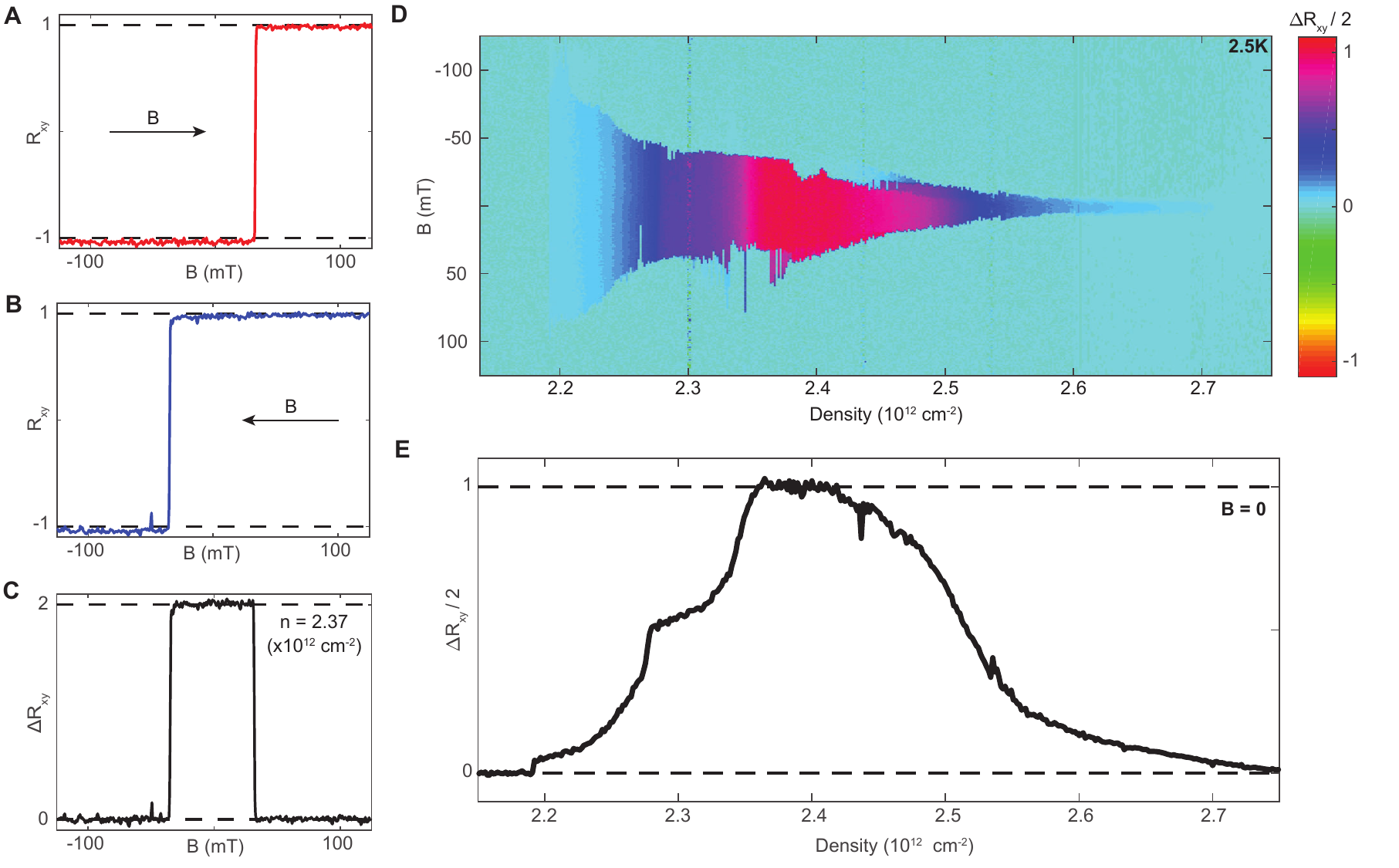}
\caption{\textbf{Exploration of magnetic order as a function of density} (\textbf{A}) Measurement of $R_{xy}$ as a function of magnetic field B as it is swept from negative to positive values. (\textbf{B}) Measurement of $R_{xy}$ as a function of B as it is swept from positive to negative values. Differences between \textbf{A} and \textbf{B} indicate the presence of ferromagnetic order in the sample. (\textbf{C}) \textbf{B} $-$ \textbf{A} as a function of magnetic field.  This value, $\Delta R_{xy}$, is nonzero below the coercive field when ferromagnetic order is present. It is equal to 2 when $ R_{xy}$ is well-quantized. \textbf{D} $\Delta R_{xy}/2$ as a function of density and magnetic field.  At a density of $2.16\times 10^{12}\text{ cm}^{-2}$, corresponding to 71\% filling of the band or $\nu=2.84$, $\Delta R_{xy}/2$ first deviates from 0 as ferromagnetic order sets in.  Ferromagnetism vanishes at $2.84\times 10^{12}\text{ cm}^{-2}$, corresponding to 92\% filling of the band or $\nu=3.68$.}
\label{fig:extentofhysteresis}
\end{figure*}

 The presence of ferromagnetic ordering in our device manifests as hysteresis in $R_{xy}$ as a function of $B$. Hysteresis implies a mismatch between measured resistances at a given $B$ for opposite signs of field training.  In other words, in a hysteretic regime, $R_{xy}(B_-)$, the measurement with negative field training (Fig.~\ref{fig:extentofhysteresis}A), differs from $R_{xy}(B_+)$, the measurement with positive field training (Fig.~\ref{fig:extentofhysteresis}B), and their difference $\Delta R_{xy}(B)$ is nonzero. At densities exhibiting a robust, well-quantized QAH state, $\Delta R_{xy} = 2h/e^2$, as shown in Fig.~\ref{fig:extentofhysteresis}C.  The presence of ferromagnetic order, the coercive fields, and the quantization of $R_{xy}$ can simultaneously be visualized by plotting $\Delta R_{xy} / 2$ as a function of the density $n$ (Fig.~\ref{fig:extentofhysteresis}D). Notably, the extent of ferromagnetism as a function of density is significantly greater than that of the QAH effect. Ferromagnetism occurs at a minimum density of $2.19\times 10^{12}\text{ cm}^{-2}$, which corresponds to $\nu = 2.84$.  Intererestingly, the coercive field of the ferromagnetic phase in tBLG is largest for the lowest values of $n$ for which hysteresis in $R_{xy}$ is detectable, terminating abruptly as the density is lowered further. As the density is increased, the coercive field decreases more or less monotonically until ferromagnetism vanishes at $2.84\times 10^{12}\text{ cm}^{-2}$, which corresponds to $\nu = 3.68$.  A quantized anomalous QAH effect is observed in a density range of $2.36 - 2.42 \times 10^{12}\text{ cm}^{-2}$ (Fig.~\ref{fig:extentofhysteresis}E).

\subsubsection*{Repeatability of hysteresis loops}

\begin{figure*}[ht!]
\includegraphics[width=4.75 in]{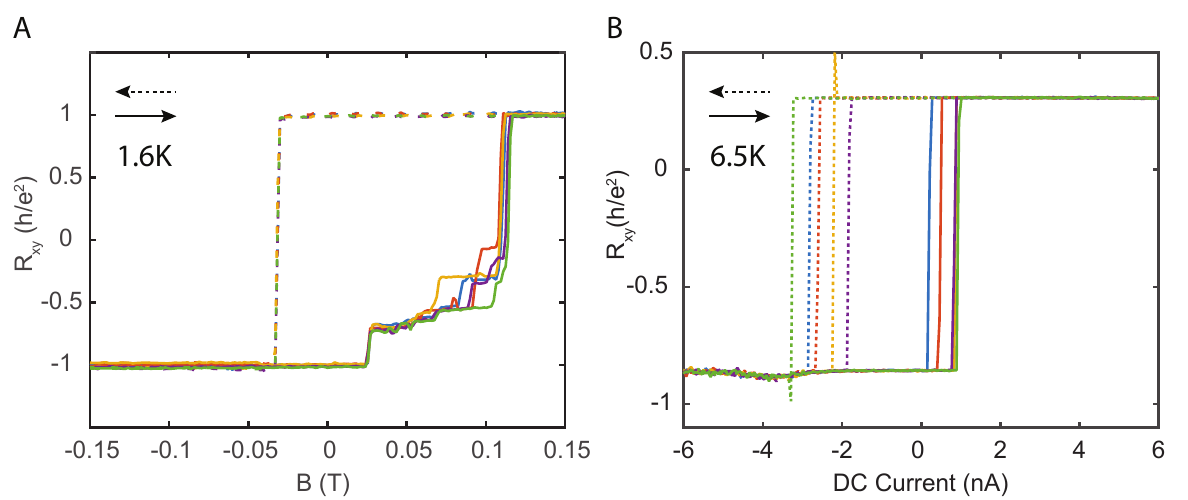}
\caption{\textbf{Repeated hysteresis loops.}
Hall resistance repeatability as a function of both applied magnetic field (\textbf{A}), and DC current (\textbf{B}) at a constant density of $2.37\times10^{-12}\text{ cm}^{-2}$. Measurements are repeated 5 times in immediate succession. The raw data is presented without any form of symmetrization. Increasing magnetic fields and DC currents are plotted as solid lines, whereas decreasing fields and currents are plotted as dashed lines. Magnetic field measurements are performed at 1.6K, whereas the DC current measurements were performed at $6.5$~K.
}
\label{fig:Repeat}
\end{figure*}

We check the repeatability of both magnetic field and DC current dependence of transport. Both measurements are presented in Fig.\ref{fig:Repeat}, in which we sweep the applied control parameter from the negative to positive value and back five times while maintaining a constant density of $2.37\times10^{-12}\text{ cm}^{-2}$.

We measure the magnetic field dependence of the Hall resistance as we sweep the field to $\pm150$ mT at $1.6$~K in Fig.~\ref{fig:Repeat}A. The structure is consistent between sweeps with many of the intermediate jumps appearing highly repeatable. Domain switching and coercive field are asymmetric in field. Structure within Fig.~\ref{fig:Repeat}A is likely related to the presence of several ferromagnetic domains with different coercive fields.  These can plausibly be explained by the presence of disorder in strain or twist angle.  DC current measurements are made with an applied current within a range of $\pm6$nA at a temperature of $6.5$~K in Fig.~\ref{fig:Repeat}B. We find that current switching is only repeatable at higher temperatures. Above $\sim6$~K, magnetic domains with smaller Curie temperatures are not magnetized, so the hysteretic behaviors with respect to both B and I are characterized by single steps in $R_{xy}$ of order $O(\frac{h}{e^2})$ (as opposed to many smaller steps). We presume that current-switching phenomena are most repeatable when the sample is left with a single ferromagnetic domain because the sign of the current-domain coupling can vary from domain to domain.

\begin{figure*}[ht!]
\includegraphics[width=2.25 in]{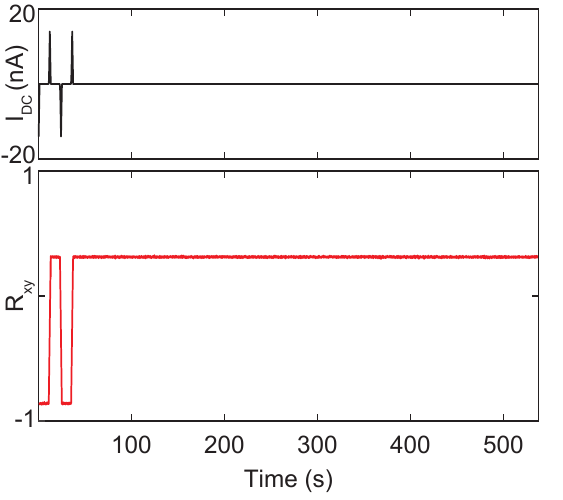}
\caption{\textbf{Magnetic bit stability.}
After a DC current pulse train switches the magnetization of the QAH magnetization several times, the bit is monitored for 500 seconds.  No measurable decay in the magnetization of the sample is detected in that time.  }
\label{fig:IStability}
\end{figure*}

 Repeatable switching of the magnetization state of the device using small pulses of DC current was demonstrated in Figure \ref{fig:3}B.  Storage of information in magnetization states is preferable to storage of information in charge states primarily because metastable magnetization states tend to be less volatile than metastable charge states.  In Figure \ref{fig:IStability} we demonstrate repeated switching of the magnetization state of the device followed immediately by a stability test of the magnetization.  Over a period of 500 seconds, the magnetization of the device does not switch or decay to within the precision of the Hall resistance measurement.

\clearpage
\subsection{Comparison of energy scales with data from magnetically doped topological insulators}
\begin{figure*}[ht]
\includegraphics[width=4.75 in]{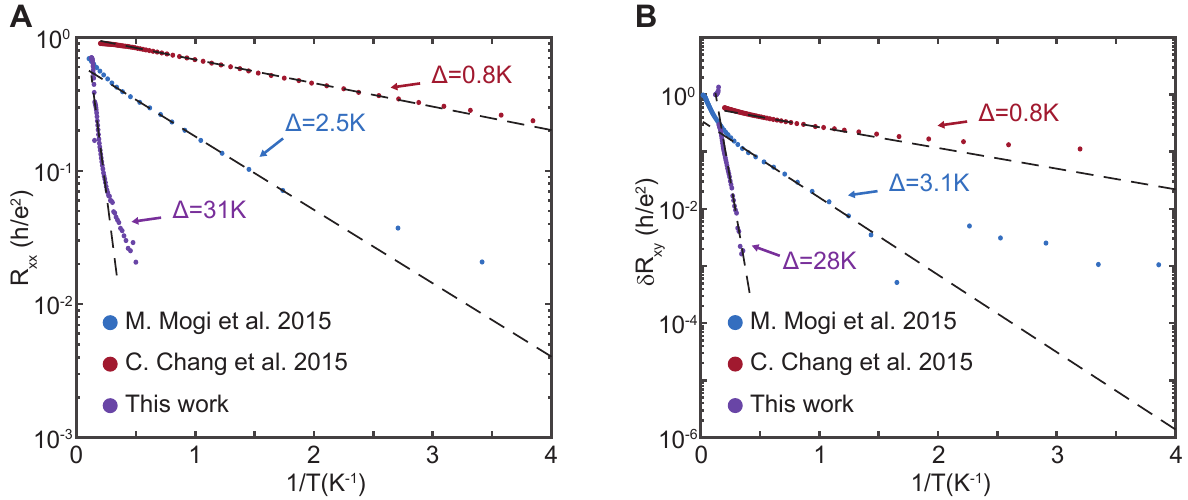}
\caption{\textbf{Zero field thermal activation gap of QAH effect in tBLG and magnetically doped TIs.} We determine the $\sigma_{xx}$ activation gap of Cr modulation doped\cite{mogi_magnetic_2015} (blue) and V doped\cite{chang_high-precision_2015} (red) $(\text{Bi,Sb})_2\text{Te}_3$ with the formulas outlined in S\ref{sec:test}. Their thermal activation data is plotted with tBLG data on the same axes. In (\textbf{A}), we fit the thermal activation of $R_{xx}$ and in (\textbf{B}) we fit the thermal activation of $R_{xy}$.
}
\label{fig:GapCompare}
\end{figure*}

Here we compare the zero field activation gap of the QAH effect in twisted bilayer graphene to those of Cr modulation doped $(\text{Bi,Sb})_2\text{Te}_3$ and V doped $(\text{Bi,Sb})_2\text{Te}_3$ ~\cite{mogi_magnetic_2015, chang_high-precision_2015}. Activation gaps have typically not been reported in these compounds. However, digitizing the published data and replotting it on an Arhennius scale shows that is well described by a single activation scale.
Applying the same fitting procedures to the data of Mogi\cite{mogi_magnetic_2015} and Chang \cite{chang_high-precision_2015}, we find gaps of around $2.8$~K for Cr doped $(\text{Bi,Sb})_2\text{Te}_3$, and $0.8$~K for V doped $(\text{Bi,Sb})_2\text{Te}_3$, as presented in figure \ref{fig:GapCompare}. The gap size for tBLG is $27$~K, which represents almost an order of magnitude improvement over the next most robust QAH sample. A comparison of the QAH gap to the Curie temperature, $\Delta_H/T_C$, reveals the difference in nature of the QAH effect in these systems. In Cr doped and V doped topological insulators, the Curie temperatures are $25$~K and $40$~K, corresponding to gap-to-Curie-temperature ratios of 0.1 and 0.02 respectively. The intrinsic magnetism in tBLG results in a $T_H/T_C \approx 3$, indicating that either intrinsic magnetism or a lack of doping induced disorder strongly enhances the robustness of the QAH effect.

\newpage
\clearpage
\subsection{Current Coupling to Domains}

The mechanism by which the current is interacting with ferromagnetic domains is poorly understood. The observed behavior is not consistent with joule heating-driven thermal relaxation of magnetic order because the application of a DC current stabilizes a magnetization state in an opposing magnetic field (Fig. \ref{fig:3}C).  This statement holds true even in the limit of extreme device asymmetry.

In this section, we consider a simple mean field model for the coupling of an applied current to the QAH order parameter, i.e. the Ising magnetization, whose sign determines the Chern number and therefore the Hall conductivity.  In particular, we show that the non-equilibrium distribution of electrons in the current-carrying state is different in the two domains, owing to a difference in population of the chiral edge states and to edge asymmetries.  As a result, the free energy of a domain contains a term which is odd in the order parameter and in the current, thus preferentially favoring a particular sign of the Hall conductivity, which can be switched by changing the sign of the current.  The effect is linear in the length of the edges.

We consider a Hall bar with translational invariance along $x$ and a confining potential along $y$.  As a simplified model we take a free electron gas with two spin states, and presume that the up spins see a positive (orbital) magnetic field $B$ and the down spins see a negative magnetic field $-B$.  The total density of electrons is such that it would fill one Landau level if all spins were polarized; this mimics the situation in a Chern insulator with $\nu=\pm 1$.  We add a repulsive interaction between species, $U n_\uparrow n_\downarrow$.  Presumably this leads to spontaneous polarization, with the Ising order parameter $\Phi= n_\uparrow -n_\downarrow$.  Let us consider the mean field theory in which without domain walls $\Phi$ becomes a constant, and all spins are polarized.  Then we have two possible states related by time reversal symmetry.

Consider a single domain.  We suppose the minority polarization states are exchange split above the Fermi energy, and model only the single spin polarization selected by the order parameter.  This is then just the usual integer quantum hall effect.  The twist is that we want to compute the free energy in the presence of a current, and then observe how it depends upon the magnetization state of the domain.  For a single domain in the Landau gauge the lowest Landau level states are specified by a single momentum quantum number $k=k_x$.  They are localized at the position $y(k)= k \ell^2 {\rm sign}(B)$, where $\ell = \sqrt{\phi_0/B}$ is the magnetic length.  Their dispersion $\epsilon(k)$ forms a flattened parabola which approximates the constant Landau level energy far from the boundaries, but rises due to the confining potential when $y(k)$ reaches either boundary.  The precise way in which it rises is non-universal and depends upon the shape of the boundaries.  Importantly, the final result will be expressed entirely in terms of properties of the two edges near the Fermi energy.  Thus we expect that the various assumptions in the model are not important, and the results are generic and model independent for a QAH state.

In this formulation, we do not quite separate the two boundaries into distinct channels, because by treating both edges as part of one whole, we automatically achieve the cancellation of unphysical effects of states far from the Fermi energy which must otherwise be put in manually.   The logic we will take is to treat the system in a sort of generalized Gibbs ensemble, in which we assume quasi-equilibrium with Lagrange multipliers for conserved and approximately conserved quantities.  The former is the charge, i.e. electron number, and the latter is the current, which is conserved when we assume the two edges are decoupled.  So with this replacement we have
\begin{equation}
  \label{eq:1}
  H \rightarrow H - \mu N - \tilde{\mu} I = \sum_k \left( \epsilon(k) - \mu - \tilde{\mu}\epsilon'(k)\right) c_k^\dagger c_k^{\vphantom\dagger}.
\end{equation}
We can from this write the free energy
\begin{equation}
  \label{eq:2}
  F/L = \int \! \frac{dk}{2\pi}\, f(\epsilon_k - \mu - \tilde{\mu} \epsilon'_k),
\end{equation}
with
\begin{equation}
  \label{eq:3}
  f(\epsilon) = - k_B T \ln \left( 1 + e^{-\beta\epsilon}\right).
\end{equation}
The current is given by
\begin{equation}
  \label{eq:curr}
  I = -\int \! \frac{dk}{2\pi}\, e \epsilon'_k n_F(\epsilon_k - \mu - \tilde{\mu} \epsilon'_k).
\end{equation}
Here the Fermi function
\begin{equation}
  \label{eq:ff}
  n_F(\epsilon) = \frac{1}{e^{\beta\epsilon}+1} = f'(\epsilon).
\end{equation}
This implies
\begin{equation}
  \label{eq:II}
  I = e\frac{\partial (F/L)}{\partial \tilde\mu}.
\end{equation}
Eq.~(\ref{eq:II}) fixes $\tilde\mu$ as a function of current $I$, and thereby, inserting this into Eq.~(\ref{eq:2}), obtain the free energy in terms of current.   We carry this out in a Taylor series in $\tilde\mu$ and $I$.

The expansion of the free energy is
\begin{align}
  \label{eq:4}
  F/L & =  \mathcal{F}_0  +\frac{1}{2} \mathcal{F}_2 \tilde{\mu}^2 + \frac{1}{6} \mathcal{F}_3\tilde{\mu}^3+ O(\tilde{\mu}^4).
\end{align}
with $\varepsilon_k =\epsilon_k-\mu$.  The first term $\mathcal{F}_0$ is a constant.  The remaining coefficients are
\begin{align}
  \label{eq:6}
  \mathcal{F}_2 &  =  \int \! \frac{dk}{2\pi}\, (\varepsilon'_k)^2 n_F'(\varepsilon_k) \approx   \frac{1}{2\pi} \left( |v_1|+|v_2|\right), \nonumber \\
  \mathcal{F}_3 & = \int \! \frac{dk}{2\pi}\, (\varepsilon'_k)^3 n_F''(\varepsilon_k) \approx - \frac{1}{\pi} \left( \frac{{\rm sign}(v_1)}{m_1} + \frac{{\rm sign}(v_2)}{m_2}\right).
\end{align}
Here $v_i = \epsilon'(k_i)$ is the velocity at the end $i$ (top or bottom of the Hall bar), and $m_i$ is the (inverse) curvature at $k_i$.  The approximation signs indicate the leading terms in the $T\rightarrow 0$ limit, i.e. for temperatures well below the AQHE gap.

With these equations we can solve for $\tilde\mu$ in terms of the $I$ up to second order in current.  Reinserting this into the formula for the free energy we obtain consistently up to third order the result
\begin{equation}
  \label{eq:10}
  F/L = \mathcal{F}_0 + \frac{1}{2e^2\mathcal{F}_2}I^2 - \frac{\mathcal{F}_3}{3e^3\mathcal{F}_2^3} I^3+ O (I^4).
\end{equation}
Finally we have obtained a term (proportional to $I^3$) in the energy which is odd in the current.  The coefficient $\mathcal{F}_3$ is odd under time-reversal, i.e. changes sign in the two domains.  So the cubic term favors one domain over the other, depending upon the sign of the current.

\subsubsection{Estimates of effect magnitude in tBLG}

To make an estimate of the magnitude of these effects, suppose $|v_1|\equiv v\gg |v_2|$ and $1/m_1 \equiv 1/m \gg 1/m_2$.  Then the cubic term in the free energy is
\begin{equation}
  \label{eq:Fcube}
  F \sim \frac{(2\pi)^3}{3\pi} \frac{\hbar^2{\rm sign}(v)}{m e^3 v^3} L I^3.
\end{equation}
Here we restored the dependence on $\hbar$.  The contribution the current to the free energy is enhanced by decreases in the edge mass and velocity, which are determined by non-universal edge physics. The free energy is particularly sensitive to $v$ and $I$, since both appear cubed, which renders making precise estimates difficult. Nonetheless, to show consistency, we take $v=5\times 10^4$m/s (a typical literature value for flat band TBG), and $m=m_e$, i.e. a unit effective mass, and a current of $I=100$nA, which is the order of the switching currents at low temperature (since the theory has been carried at $T=0$). This gives an energy $F=4.0$meV, which is similar to the magnetostatic energy assuming an orbital moment per electron of a few Bohr magnetons.  Uncertainties in the edge properties as well as thermal renormalizations not taken into account here make it hard to make a more quantitative comparison at present.  These are interesting subjects for future work.

\end{document}